**Title:** The COVID-19 Pandemic, Community Mobility and the Effectiveness of Non-pharmaceutical Interventions: The United States of America, February to May 2020

Running title: Impact of COVID-19 interventions in the U.S.

**Authors and affiliations:**

Ian E. Fellows[1], Rachel B. Slayton[2] and Avi J. Hakim[2]

[1] Fellows Statistics, San Diego, US

[2] US Centers for Disease Control and Prevention, COVID-19 Response Team, Atlanta, US

**Corresponding author:**

Ian E. Fellows

1600 Clifton Rd, NE

US1-2

Atlanta, GA 30329

619.495.6688

Ymx4@cdc.gov

**Summary:** We found a strong relationship between reproductive number and community mobility. Mobility changes in states that did not implement particular non-pharmaceutical interventions suggests that the decrease in mobility associated with time may be due to individual level mitigation measures.

[Initial draft: 4/30/2020; Completed CDC clearance and peer review: 6/30/2020]

# Abstract

Background: The impact of individual non-pharmaceutical interventions (NPI) such as state-wide stay-at-home orders, school closures and gathering size limitations, on the COVID-19 epidemic is unknown. Understanding the impact that above listed NPI have on disease transmission is critical for policy makers, particularly as case counts increase again in some areas.

Methods: Using a Bayesian framework, we reconstructed the incidence and time-varying reproductive number ($R_t$) curves to investigate the relationship between $R_t$, individual mobility as measured by Google Community Mobility Reports, and NPI.

Results: We found a strong relationship between reproductive number and mobility, with each 10% drop in mobility being associated with an expected 10.2% reduction in $R_t$ compared to baseline. The effects of limitations on the size of gatherings, school and business closures, and stay-at-home orders were dominated by the trend over time, which was associated with a 48% decrease in the reproductive number, adjusting for the NPI.

Conclusions: We found that the decrease in mobility associated with time may be due to individuals changing their behavior in response to perceived risk or external factors.

BACKGROUND

With no vaccine and few therapeutics currently available to respond to the COVID-19 pandemic, countries have implemented a variety of non-pharmaceutical interventions (NPI) aimed at mitigating the virus' spread. However, there is substantial uncertainty regarding the effectiveness of these measures.

To assess the epidemiological effect that any intervention has on the spread of the virus through a population, it is critical to evaluate its impact on the shape of the incidence curve and the effective reproductive rate (Rt), which characterizes the epidemic growth. In an uncontrolled epidemic ($Rt>1$), additional NPI may be needed, while in a controlled epidemic ($Rt \leq 1$), some NPI can be relaxed.

Efforts to evaluate the impact of Coronavirus disease 2019 (COVID-19) NPI have found an association between the use of NPI and community mobility in the US (1) and in Spain between statewide stay-at-home orders and disease spread across provinces (2). Others have highlighted the additive value of NPI but there is little other research estimating the effect of individual NPI (3). A model of 11 European countries found that NPI were associated with a reduction in Rt, with national stay-at-home orders having a particularly large impact (4).

COVID-19 epidemic models without adaptations to standard epidemiological methods are difficult to accurately construct when there is variable testing capacity. While case counts can provide nearly up to date information on disease trends, they are confounded by changes in testing rates, prevalence, and how efficiently individuals are targeted for testing. An increase in cases could be caused by a true increase in incidence or by expanded testing. However, as access to tests has been increasing over time as it generally has in the United States (See appendix), then declines in daily case counts indicate a decrease in incidence. This relationship allows us to use cases to enforce an upper bound on incidence increases and improve model precision. We used case and death counts to develop a novel Bayesian model to assess the impact of gathering limitations, school and business closures, and stay-at-home orders on the COVID-19 epidemic in the US.

METHODS

Our model uses Bayesian methods to infer COVID-19 incidence and time-varying reproductive number in the US from case and death counts. The model was applied to the country and disaggregated to the state level. We inferred the incidence pattern to represent a "Nowcast" of the pandemic (i.e., cases that have occurred but have not yet been reported due to inherent reporting lags) and can be used to provide daily insight into the rate of disease transmission over time. The heterogeneity of the United States' subnational approach to combatting COVID-19 allows us to examine how effective selected NPI are by looking at changes in reproductive number occurring after application of an intervention or combination of interventions (5).

Death and case counts by day in the US were curated by the COVID Tracking Project going back to January 22, 2020 (6). Google's Community Mobility Reports provide information on the percentage change in visits and length of stay to various types of locations compared to a baseline or median level of activity on that day of the week from January 3 to February 6, 2020 (7). Location categories include 'Grocery and pharmacy,' 'Parks,' 'Transit stations,' 'Retail and recreation,' 'Residential,' and 'Workplaces' (7). These data come from individuals whose Location History setting is turned on in their Google account. The percentage of the population constituted by this group is unknown. To construct an overall mobility score, we performed a principal components analysis (PCA) excluding 'Parks' and 'Grocery and pharmacy' to find the best linear combination of metrics desciribing the mobility trends. The 'Parks; metric excluded because it posed limited transmission risk and the 'Grocery and pharmacy' metric was excluded because it includes essential activities. The PCA parameters were used to generate a weighted average (8). We produced a a 7-day moving average to account for the periodicity of mobility patterns.

There are several key challenges to overcome when inferring COVID-19 incidence curves. First, both death and case counts are delayed from time of infection. Secondly, while the lag time for case counts is considerably shorter than death counts, case counts are difficult to rely on in practice because the number of cases diagnosed on any day is a function of the number of infected individuals available to detect, the amount of testing capacity available to diagnose them, and how effectively those tests are deployed. This difficulty has lead other authors evaluating NPI to solely focus on death counts (6).

However, if one assumes that the probability an infected individual gets diagnosed increases or stays the same over time, percentage changes in the true incidence are bounded above by changes in the case incidence (i.e., the expected number of newly infected individuals on a day who will go on to be diagnosed). Thus, an increase in cases may be due to increased testing but a sustained decrease in cases signals a true decrease in incidence. This assumption is reasonable in the United States as the number of tests conducted has increased over time and the percentage of tests that are positive has generally decreased or remained constant over the study period (9, 10). A detailed description of the modeling approach can be found in the appendix.

We combined the lag-time distributions for cases and deaths with the observed counts in a Bayesian model which respects the bounding constraint discussed above. The primary output of this model is a probability distribution for the death incidence, defined as the number of individuals infected on a day who will go on to die. Provided the infection fatality rate is constant, this quantity is proportional to the true incidence. From the incidence distribution, we can then calculate the time varying reproductive rate (Rt) (Equation 4.1 of (11)).

Generalized Additive Models for NPI evaluation

We evaluated the effects of NPI by taking the estimated log Rt curve for each state and fitting a generalized additive model (GAM) with random effects (12). This model includes a smoothed non-parameteric term for date common across all states, and fixed effects that are zero before the NPI implemented and one after for each of the NPI evaluated. Partial and full stay-at-home orders were combined as this provided a better fit and there were few mandatory partial stay-at-home orders (5). Each of the fixed effects captures the effects of one of the interventions. The smoothed date term captures any changes in mitigating behavior not triggered by an NPI or overall changes in viral contagiousness (for example due to changes in weather). A random effect was also included at the state level, as there was considerable variability in initial Rt between states. Uncertainty intervals were calculated by bootstrap resampling at the state level. The bootstrap was implemented such that, each state is drawn with replacement and given a unique state identifier label such that if a state is drawn twice, each draw will have a unique state label. The GAM model is then fit to each of the bootstrap samples, and the distribution of these fit parameter estimates are used to construct the bootstrap confidence intervals. Mobility was not included in the model because we assumed NPI effects were mediated by changes in mobility.

NPI effects on mobility were evaluated using a similar GAM except that the outcome was mobility score and terms for day of the week were also included. Only dates prior to April 21, 2020 were included in the NPI analysis so as to not include areas of the Rt curve with high posterior uncertainty and exclude any reopening effects. Confidence intervals were constructed via bootstrap resampling of states. Five NPI measures by state were extracted from official documents and announcements, and media reports: 1) gathering size limitation- state ordered limitation on the size of groups or events that can be conducted, except those enacted as part of a full stay-at-home order; 2) school closure state ordered closure of some or all in-person K-12 educational institutions, except those enacted as part of a full stay-at-home order; 3) business closure- a state level order closing some or all businesses, except those enacted as part of a full stay-at-home order, this could include all services deemed non-essential, but often applies to a select subset such as gyms or barbers; 4) partial stay-at-home order: A stay-at-home order applied to a subset of the state's population, this could apply to a geographic area or a category of individuals such as those at risk for severe outcomes; 5) full stay-at-home order: A stay-at-home order applied to the entire state population, potentially excluding essential personnel (13-15).

RESULTS

We estimate that on March 29, 2020, 2,673 transmission events of SARS-CoV-2 occurred that would later result in an associated death (95% Probability Interval (PI) 2,329, 3,078). One month later on April 29th, 2020, this number decreased to 906 (95% PI: 272, 1824), with an estimated reproductive number (Rt) of 0.70 (95% PI: 0.40, 1.04).

Figure 1 shows the national estimated reproductive number and reported and estimated incidence of fatal infections over time aggregated to the whole country. Initially, the reproductive number of the virus was high and in this initial phase it was possible to calculate an estimate of the unmitigated reproductive number (R0). On March 14, the reproductive number was estimated to be 2.89 driven largely by the outbreak in New York where we estimate approximately half of new infections nationally were located. Using March 14, 2020, as a baseline date for unmitigated transmission and only including the 32 states that had reported at least 5 cases and whose mobility score decreased by less than 10%, we estimate the average R0 across states to be 2.37 (95% PI: 2.22, 2.52). Estimates of reproductive number by state at baseline are shown in Figure 1 of the appendix.

Over the course of the next few weeks the reproductive number decreased dramatically. Through most of April, the Rt hovered just below 1. In May, there may have been a further reduction in Rt, but uncertainty in the curve makes any conclusion premature as an Rt above 1 is within the 90% probability interval.

There was considerable heterogeneity in Rt by state after the dramatic Rt reduction, with Appendix Figure 2 showing a snapshot of Rt six weeks after baseline (April 25, 2020). Many states are near or below one, indicating epidemic control; however, several still show a trend toward increasing infection rates (Rt > 1). Incidence curves for each state can be found in the appendix.

Stay-at-home orders (either partial or full) were estimated to have decreased Rt by 13% (95% confidence interval (CI): 3%, 22%) when implemented (Figure 2). School closures showed a smaller effect with a 5% reduction (95% CI: -1%, 11%). While school closures, gathering limitations and business closures trended toward decreasing Rt, they did not show individual effects large enough to conclude that they reduce Rt. Their effects were estimated to be 5% (95% CI: -1%, 11%), 5% (95% CI: -4%, 12%), and 2% (95% CI: -8%, 11%) respectively. If all four interventions were enacted, we estimated a 22% reduction in Rt.

The largest effect in the model was the time trend, which, controlling for mitigation measures by state, showed a reduction of 48% (95% CI: 40%, 56%) from March 15, 2020 to April 20, 2020.

If mobility mediates the relationship between NPI implementation and reproductive number it would be expected that the NPI would decrease both mobility and Rt. We found a modest relationship between use of NPI and reductions in mobility. While overall mobility declined 20%-60% across states, the reduction that could be attributed to NPI was 4% (95% CI: 2%, 6%) for stay-at-home orders, 2% (95% CI: 1%, 4%) for business closures, 3% (95% CI: 1%, 6%) for gathering limitations and 2% (95% CI: 0%, 5%) for school closures.

Figure 3 displays the relationship between time, reproductive number and mobility by state. Both Rt and mobility are relatively stable prior to March 15. Then both decrease by 40%-50%, reaching a new stable point near April 1. The reduction in mobility during this period cannot be explained by stay-at-home orders alone, as similar change patterns are seen in states that never implemented stay at home orders.

We found an 85% (95% CI: 82%, 89%) correlation between Rt and mobility (with periodicity removed by applying a 7-day moving average)). If every percentage increase activity had an equal chance of causing a transmission event, we would expect there to be an approximately one-to-one relationship between reductions in mobility and reductions in reproductive numbrt. Fitting a line to the data, we found a slope of 1.02 (95% CI: 0.93, 1.10) closely matching this expectation. Plots of this relationship by state are included in the appendix.

Trace plots of the association between mobility and Rt for different NPI reveal a large downward trend in mobility and Rt in states that had not yet implemented a state level intervention (Figure 4). Mobility was lower among states that enacted each NPI and Rt was modestly lower. The largest difference is seen among states that did and did not implement statewide stay-at-home orders. School closures occurred over a more compressed timetable than the other NPI, though the downward trajectories in both mobility and Rt are visible prior to implementation of assessed NPI.

**Conclusions**

School closures and stay-at-home orders were associated with Rt reduction but the large changes in mobility over time cannot be explained by the four NPI we modeled alone. External factors played a larger role as evidenced by the large national time trends. Individuals may have changed their mobility behaviour in response to perceived risk, guidance from community and faith-based organizations, employers providing opportunities to telework, city/county government actions, and media coverage of the pandemic including measures being taken by other countries and states (16, 17). Additionally, government officials and public health professionals made numerous recommendations to mitigate COVID-19, including President Trump's unveiling of a 15 day plan to slow the spread of COVID-19 (18). While data on susceptibility of SARS-COV-2 to temperature and humidity changes are inconclusive, changes in mean temperature in the contiguous United States in February and March werelikely not sufficient to explain such a sizeable reduction in incidence (19-21).

A real-time understanding of the reproductive number and the impact of mitigation measures is critical for mitigating the negative consequences of both COVID-19 and the secondary effects of NPI. Our analysis provides daily estimates of incidence and Rt through a novel approach that leverages case counts, which provide a more timely view of the epidemic, while avoiding confounding from increasing testing rates. The ability to estimate Rt on a daily basis is essential given the World Health Organization's identification of a reduction of Rt to less than 1 for at least 14 days as the key measure for identifying that the epidemic is controlled (22).

This state-level analysis of data from the US provides information that may help guide state-level mitigation efforts and a framework for monitoring the epidemic at sub-state levels. Until vaccines and therapeutics for COVID-19 become available, NPI will play a primary role in decreasing disease transmission. To be sustainable, these measures need to be feasible for people to adhere to. It is important that measures be implemented in a phased approach starting with the least disruptive and balance socioeconomic and public health impacts (23). Additionally, NPI measures that do not rely on mobility reduction, such as the wearing of face coverings and support for hand and respiratory hygiene may allow for both safe gatherings and businesses to operate (24).

Our estimates of the effectiveness of NPI are considerably lower that some other modeling studies, with stay-at-home orders being the only one that achieved a statistically significant effect. This could be due to our inclusion of a common non-linear time effect capturing external drivers (25-27). Our findings indicate that care should be taken not to overestimate the effect of NPI and further suggest the potential utility of reassessing projections of the number of deaths without NPI and counterfactual estimates of deaths averted resulting from their use.

Our findings are limited by not including the measures listed below in our model. State level data on personal hand and respiratory hygiene are unavailable. Use of cloth face coverings likely had a small effect on the model as guidance for the use of cloth face coverings was released on April 3, by which time Rt had already decreased significantly. Data on mobility come only from individuals whose Location History setting is turned on in their Google account. Other limitations include different definitions of essential businesses across states and the uneven enforcement of NPI.

Whereas other models focus solely on the use of NPI, our model also examines mobility as a proxy for physical distancing and captures non-NPI related changes with a non-linear time trend common across all states. Our findings reveal that the decline in mobility seen over the study period is highly correlated with the observed reduction in Rt, suggesting mobility's potential as a proxy for Rt. Such a proxy indicator may be especially useful for lower administrative levels (e.g., counties) that lack sufficient data to estimate Rt.

Maintaining the reductions in incidence and mortality may be feasible as stay at home orders are lifted provided individuals maintain their decreased mobility and increased physical distancing (3, 28). Our findings provide data that may help policy makers to facilitate communication with the public and set expectations. For example, if R0 is 2 and the relationship found between mobility and Rt continues to hold, a locality will on average need to reduce mobility to about a 50% of baseline to reach epidemic control. If individual action is insufficient to generate such a reduction, stay-at-home orders may be necessary.

Lifting of stay-at-home orders may be less important than how the public perceives this lifting. If the public believes there is still a risk from COVID-19 and maintains individual level precautions at similar levels, we would expect only a modest increase in the reproductive number. On the other-hand, if the public perceives the lifting as signalling the end of the danger and returns to a baseline levels of mobility and physical distancing we would expect the reproductive number to approximately double. Additional measures such as increased hand and respiratory hygiene, timely and effective contact tracing, and the wearing of cloth face coverings could mitigate some of this rebound. Community buy-in for sustaining NPI adherence may be facilitated through culturally relevant risk communications and a harm reduction approach to COVID-19 mitigation that promotes personal controls such as cloth face covering wearing, engineering controls such as barriers between cashiers and customers, and administrative controls such as restrictions on certain kinds of gatherings.

There are no easy answers or simple solutions in the response to the COVID-19 pandemic. The downstream effects of even small reductions in mobility may be substantial when they are broadly implemented. Epidemic control may be achievable without stay-at-home orders if individuals reduce their mobility. Longterm adherence to stay-at-home orders may be challenging. These results suggest that reduced COVID-19 transmission was associated with a decrease in mobilty and that the reduction in both transmission and mobility were observed in states that had no stay-at-home orders.

# Appendix

**The relationship between death and case rates**

We wish to reconstruct an incidence curve using information present in the daily death and case counts. This process is challenging because the lag time between infection and death can easily exceed four weeks (29). Diagnoses provide more current information for an epidemic curve, but are difficult to interpret due to the effects of changing testing criteria on diagnosis counts.

Let $\lambda_t$ be the expected number of individuals infected on day $t$, $\lambda_t^D$ be the expected number of those infected who will go on to die, and $\lambda_t^C$ be the expected number who will go on to become a diagnosed case. Further, let $q_t$ be the probability that a person infected at time $t$ will be diagnosed, and $r$ be the infection fatality rate. We will call $\lambda_t^C$ the case incidence and $\lambda_t^D$ the death incidence.

Assuming that the infection fatality rate is constant, the death incidence is proportional to incidence ($\lambda_t^D = \lambda_t r$). The case incidence is can be expressed as

$$\lambda_t^C = \lambda_t q_t.$$

$q_t$ is difficult to estimate based on data commonly available. For many countries it is reasonable to assume that testing capacity is increasing and thus $q_t \geq q_{t-1}$. This implies that the percentage increase in $\lambda_t$ is bounded above by the change in $\lambda_t^C$

$$\frac{\lambda_t^D}{\lambda_{t-1}^D} = \frac{\lambda_t}{\lambda_{t-1}} \leq \frac{\lambda_t^C}{\lambda_{t-1}^C} \qquad (1)$$

Note that while diagnosis probability may vary or may go down at the country or city level, the assumption around an increasing $q$ is relative to the whole state and is not necessarily invalidated by decreasing testing probabilities in sub-state localities.

**Lag-time distributions**

For our developments below, we will require the distribution of the time between infection and diagnosis and the distribution of the time between infection and death. Linton et al. (29) estimates the time from infection to onset of symptoms as a log-normal distribution with median 4.6 days and standard deviation of 3.9 days. This distribution is consistent with the incubation periods estimates reported in the meta-analysis of Khalili et al. (32), in which 8 of the 15 average incubation period estimates were between 4 and 6.

Linton et al. (29) also estimates of the time from symptom onset to death distribution as a log-normal distribution with median 17.1 and standard deviation 11.6 (29). This is similar to the average estimates of Shi et al. (18.7 days) and Ruan et al. (18.42 days) (30, 31) . However, all of these estimates come from very early in the pandemic in China. We were able to estimate a lag time estimate of 13.7 days with a standard deviation of 7.8 from 5,624 deaths that occurred in the United States between March 1st and March 31st. These distributional parameters were used in a log-normal distribution for our time from onset to death distribution.

We construct the time distribution for time from infection to death ($f^D$) as the sum of the incubation period and time from onset to death distributions. This distribution is discretized by day (i.e., normalized so that the sum over the days is one) yielding an expected value of 19.2 days. The time from infection to diagnosis distribution $f^C$ was set to be the sum of the incubation period and the time from onset to first clinical visit. Khalili et al. provide a pooled estimate of 4.82 days (32). The standard deviation for this period is not well described by the literature. For our analysis, we used survey data on time to seek care for influenza-like illnesses collected by the US CDC (33). Fitting a log-normal distribution to these percentiles yielded a distribution resulted in a log-normal distribution with mean 4.1 and standard deviation 4.5. The average time delay from infection to diagnosis in the combined distribution was then of 9.8 days.

Niether distribution incorporates reporting delays as these may vary across state and change over time. Modeling these dynamics are left to future work.

**Case/death counts as Poisson processes**

Let $N_t^D$ be the number of deaths for a locale on day $t$. We formulate the likelihood as a Poisson process, which has been a staple of epidemic curve reconstruction for over 30 years (34)

$$N_t^D \sim Poisson(d_t),$$

where

$$d_t = \sum_{i<t} \lambda_i^D f^D(t-i)$$

is the expected number of deaths at time $t$.

Similarly for cases we have

$$N_t^C \sim Poisson(c_t)$$

where

$$c_t = \sum_{i<t} \lambda_i^C f^C(t-i).$$

In order for λ to be well identified, it is usual to enforce some smoothness constraint or prior on it (35-37). Bacchetti et al. in particular proposed penalizing roughness in the log scale so as to favor the continuation of exponential growth or decline. We take a similar approach, but in a fully Bayesian framework.

Defining $s$ as the first day that a fatality occurs, we reparameterize $\lambda^C$ in terms of log differences

$$\log(\lambda_t^C) = \log(\lambda_{t-1}^C) + \alpha_t \quad \forall\, t > s - 60$$

and set the initial $\lambda^C$s to be a relatively small value chosen such that the expected first case would be approximatly at its observed value given a constant $\lambda$

$$\lambda_{s-60}^C = 0.02780049$$

This value has limited effect on all but the earliest time regions before any cases were observed.

We can then define a prior on theta that enforces smoothness

$$\alpha_t \sim Normal(\alpha_{t-1}, \sigma^C).$$

This prior puts higher probability on smooth curves that increase or decrease steadily for large parts of the observation period. This corresponds to exponential increases or decreases in terms of $\lambda^D$. σ controls the degree of smoothness and is given a half normal prior with standard deviation 0.05.

In reparameterizing $\lambda^D$ we also wish to enforce the constraint in Equation 1. To do this in a way that maintains likelihood differentiability, a soft minimum function is utilized

$$softmin(a, b, \eta) = \frac{ae^{-\eta a} + be^{-\eta b}}{e^{-\eta a} + e^{-\eta b}},$$

which is approximately equal to the minimum of $a$ and $b$ as $\eta$ increases. $\lambda^D$ is then defined as

$$\log(\lambda_t^D) = \log(\lambda_{t-1}^D) + softmin(\theta_t, \alpha_t, 10), \quad \forall\, t > s - 60$$

where 10 is chosen for $\eta$ in order to balance the trade-of between approximation of the minimum function and computational feasibility of Bayesian inference(38). Inspection of the approximation of the fit model across a number of locales indicates good agreement between the actual minimum and the softmin approximation. As it is questionable whether the constraint holds in the period prior to testing roll-out in the United States, it is only fully in effect after 100 positives have been identified at the locale and has no influence before the first positive was identified.

The initial $\lambda^D$ is set to be identical to $\lambda^C$

$$\lambda_{s-60}^D = 0.02780049$$

The prior for $\theta$ is set to

$$\theta_t \sim TruncatedNormal(\theta_{t-1}, \sigma^D, upper = \log(1.35)).$$

As with $\alpha$, this prior puts higher probability on smooth curves. $\sigma^D$ controls the degree of smoothness and is given a half normal prior with standard deviation 0.05. The prior distribution of $\theta$ is truncated at $\log(1.35)$ to prevent sharp incidence changes in locales with limited data. The particular value is chosen as there is evidence that, left unrestricted, COVID-19 cumulative death counts grow at a rate near or below 35% a day(39). Using the COVID-19 serial interval distribution in (40) and the early reproductive number formula from (11), this would imply a maximum reproductive number of 3.85, which is larger than most peer-reviewed estimates of R0 (41).

**Data processing and measurement error**

As would be expected in an evolving pandemic, case and death count data contain a number of data artifacts due to reporting errors and delays. To handle these, we first process counts through a rules-based system to smooth out improbable observations. First, if a count is over twice that of the previous day and the previous day's count was > 2000, we view this as a likely mass reporting of backlogged cases/deaths. For example, on one single day, New York state reported all of the previous probable COVID-19 related deaths that were not verified through testing. In these cases, we set the current day's count to be equal to the previous day. Otherwise, in cases where two sequential days have counts where one is twice the other and one has a count > 20, the two are averaged, and the average count is used for both. This condition is expanded to a 30% difference when the lesser count is > 5000. Negative counts (due to revisions) are subtracted from the most recent observed counts prior to the revision.

Even with the processing of these outliers, we observe significantly higher variation in both case and death counts in many countries than would be expected by a Poisson model. This increase is not well fit by a negative binomial distribution with constant dispersion, which is often used for over-dispersed counts. The degree of additional variation takes different forms depending on the locale, and varies by both level (i.e. value of $d_t$ and $c_t$) and over time as different reporting structures are put into place.

To account for the changing degree of measurement error, we employ a Gamma-Poisson model, where the rate of observed deaths and cases are Gamma distributed (42). This Gamma measurement error leads to a negative binomial distribution for the observed counts with means $(d_t, c_t)$ and time varying dispersion parameters $(\phi_t^D, \phi_t^C)$. If $O_t^D$ and $O_t^C$ are the observed numbers of deaths and cases respectively, we have

$$O_t^D \sim NegativeBinomial(d_t, \phi_t^D)$$

$$O_t^C \sim NegativeBinomial(c_t, \phi_t^C)$$

The $\phi$ values are estimated by first fitting a smoothed curve to the counts via generalized additive modeling in order to estimate $d_t$ (or $c_t$ in the case of case counts). Then the variance of $O_t$ over time is estimated by fitting another generalized additive model to the squared residuals from which estimates of $\phi_t$ can be calculated.

**Reproductive number by state**

Figures 1 and 2 show Rt estimates by state at baseline (March 14th) and 6 weeks later (April 23rd). There is considerable heterogeneity by state both at baseline and 6 weeks later. The large reduction in Rt over this time period is evident.

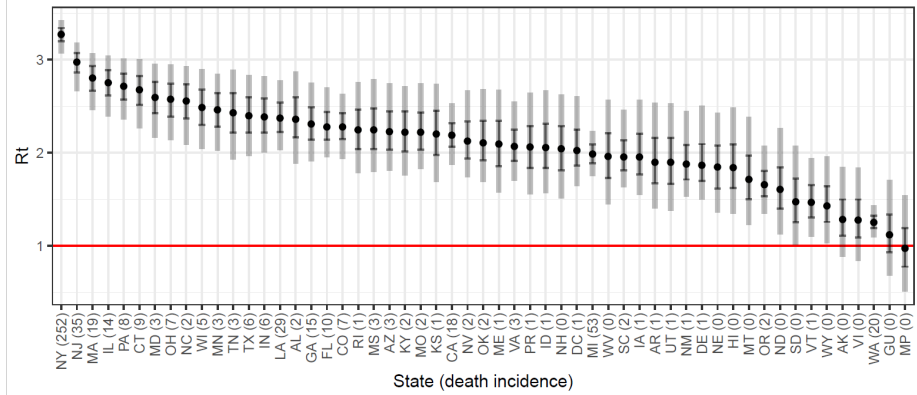

Figure 1: Estimated reproductive number by state on 2020-03-14 with IQR and 90% probability bands. The estimated number of individuals infected on this date who will go on to die is noted in parentheses.

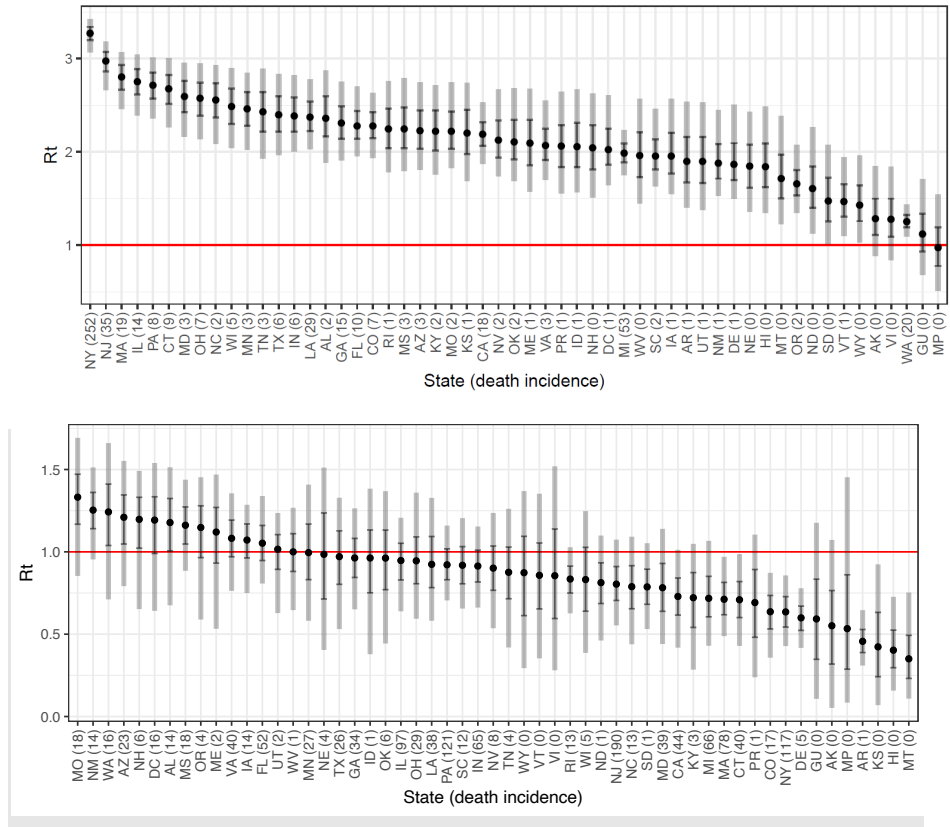

Figure 2: Estimated reproductive number by state on 2020-04-25 with IQR and 90% probability bands. The estimated number of individuals infected on this date who will go on to die is noted in parentheses.

**Plots by state**

The figures below show the death incidence, defined as the number of infections per day that will eventually lead to death. The case incidence is the number of incident infections that will be diagnosed, and is equal to the true incidence times the probability that an infection at that time will go on to be diagnosed. *E*(*Deaths*) and *E*(*Cases*) are the expected number of observed deaths and cases reconstructed from the model. These are useful in validating model fit. Additionally, the observed case and death counts are included along with the timing of various NPI measures (up to 4-24).

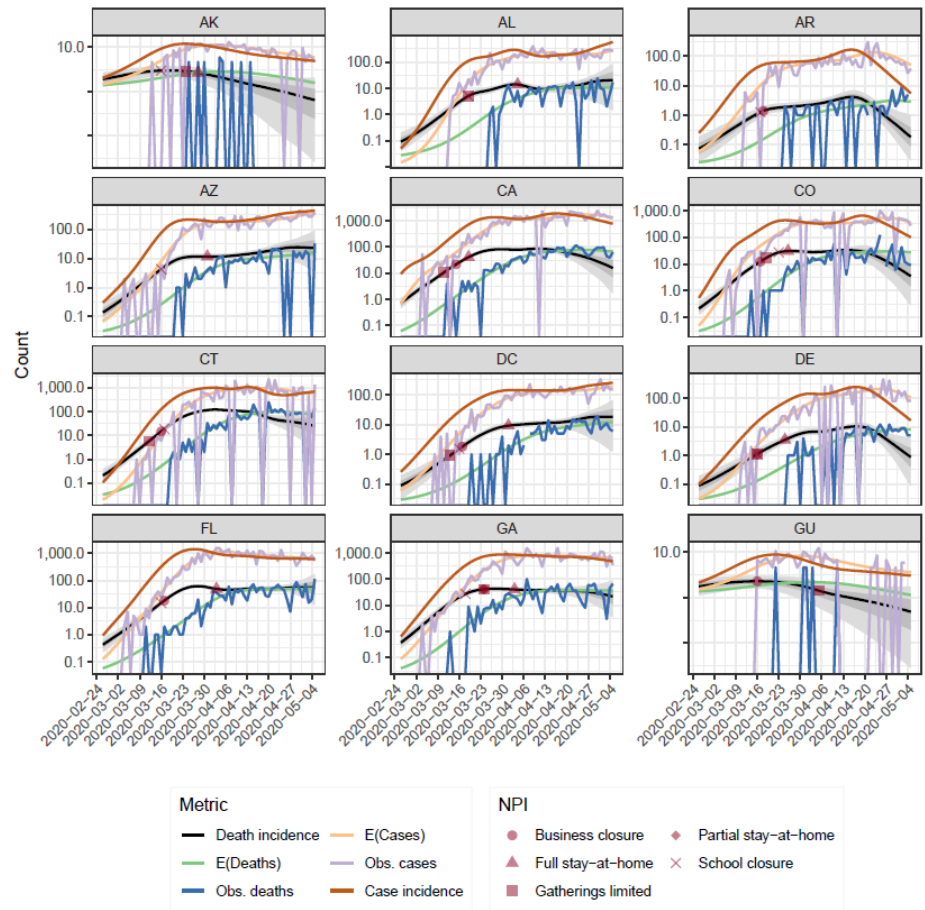

Figure 3: Death incidence plots by state

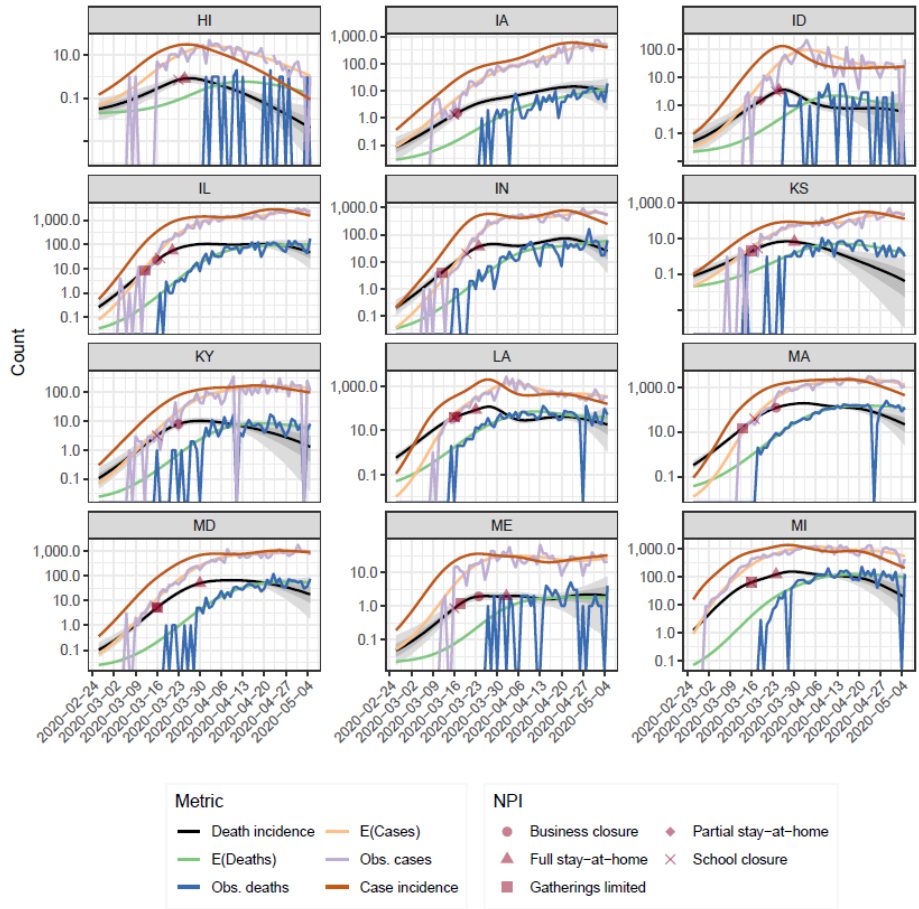

Figure 4: Death incidence plots by state

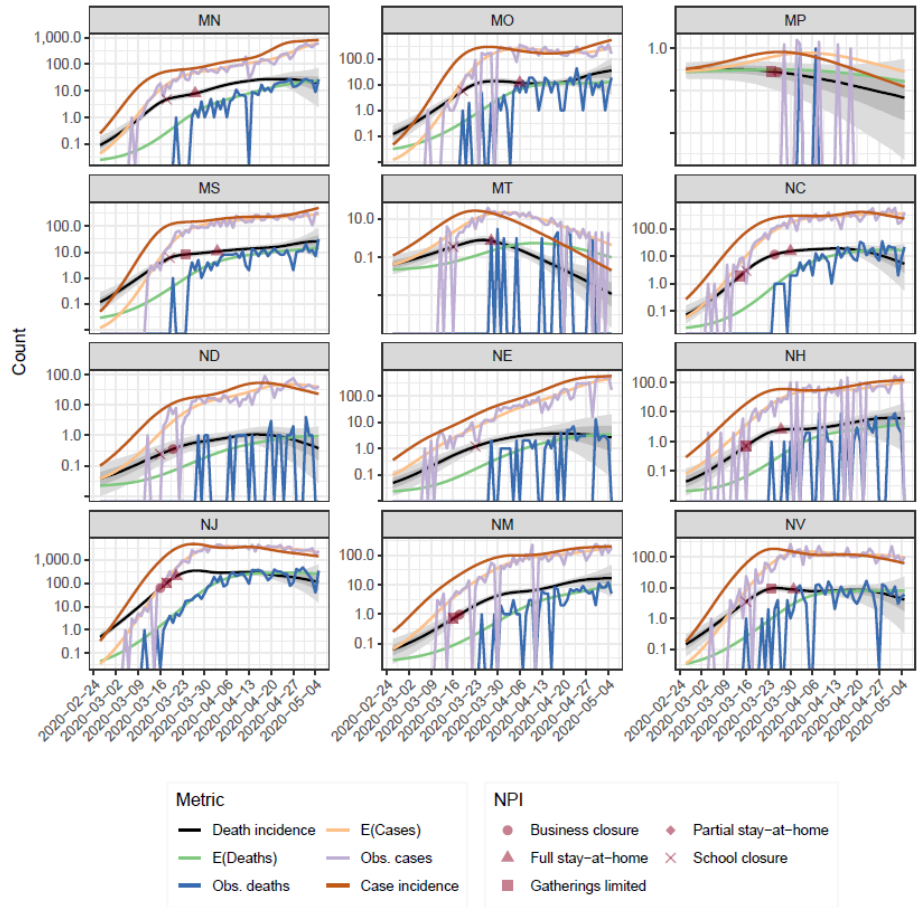

Figure 5: Death incidence plots by state

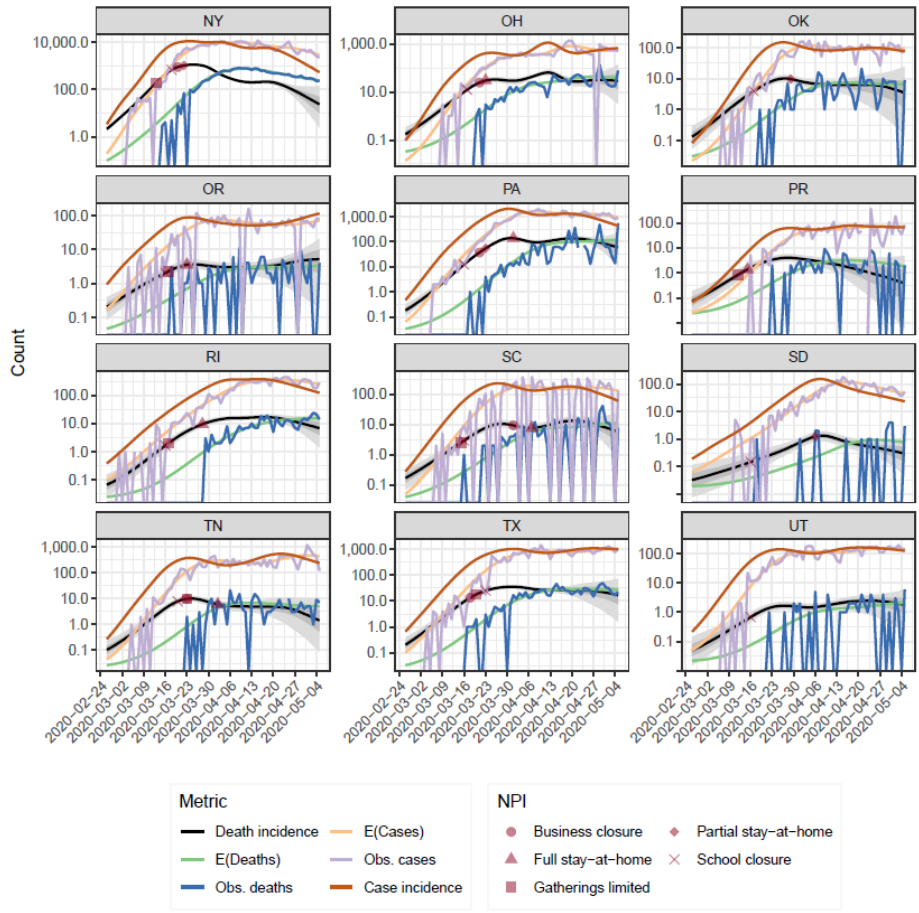

Figure 6: Death incidence plots by state

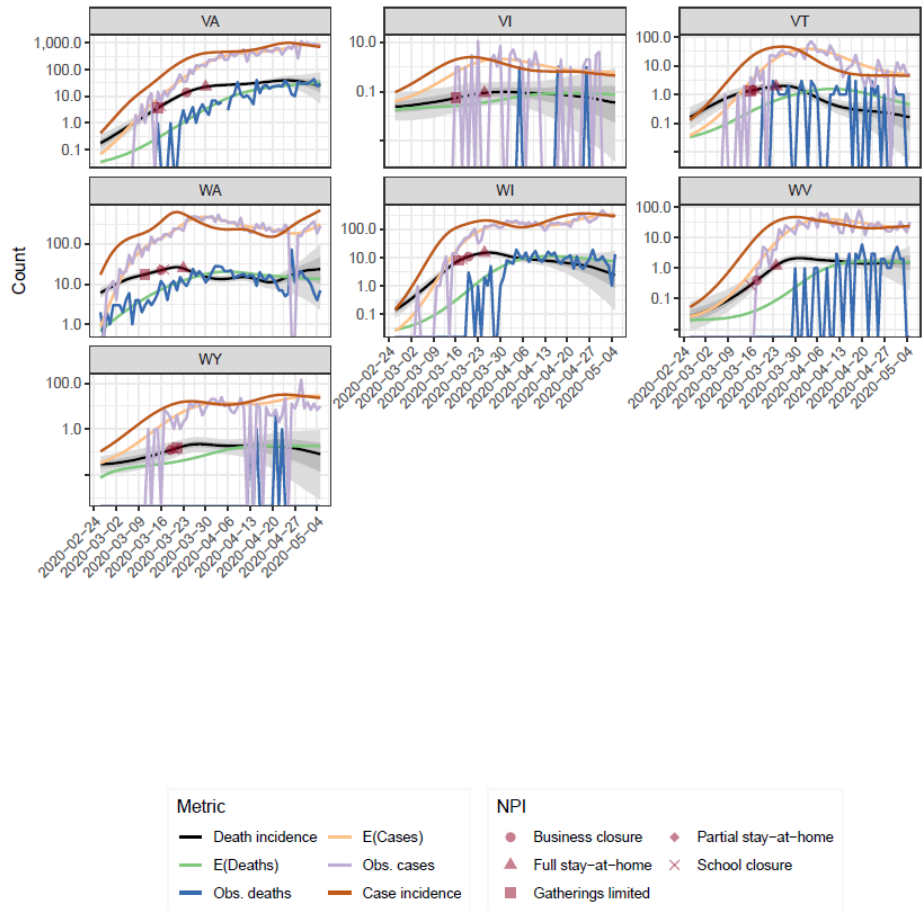

Figure 7: Death incidence plots by state

**Rt and mobility by state**

The Figure below displays the relationship between Rt and mobility by state. The y-axis represents the reduction in Rt compared to its maximal value and the x-axis is mobility. States start in the upper left, with no mobility reduction and then shift to the lower left as mobility decreases. A similar, nearly linear, reduction pattern is seen across many states.

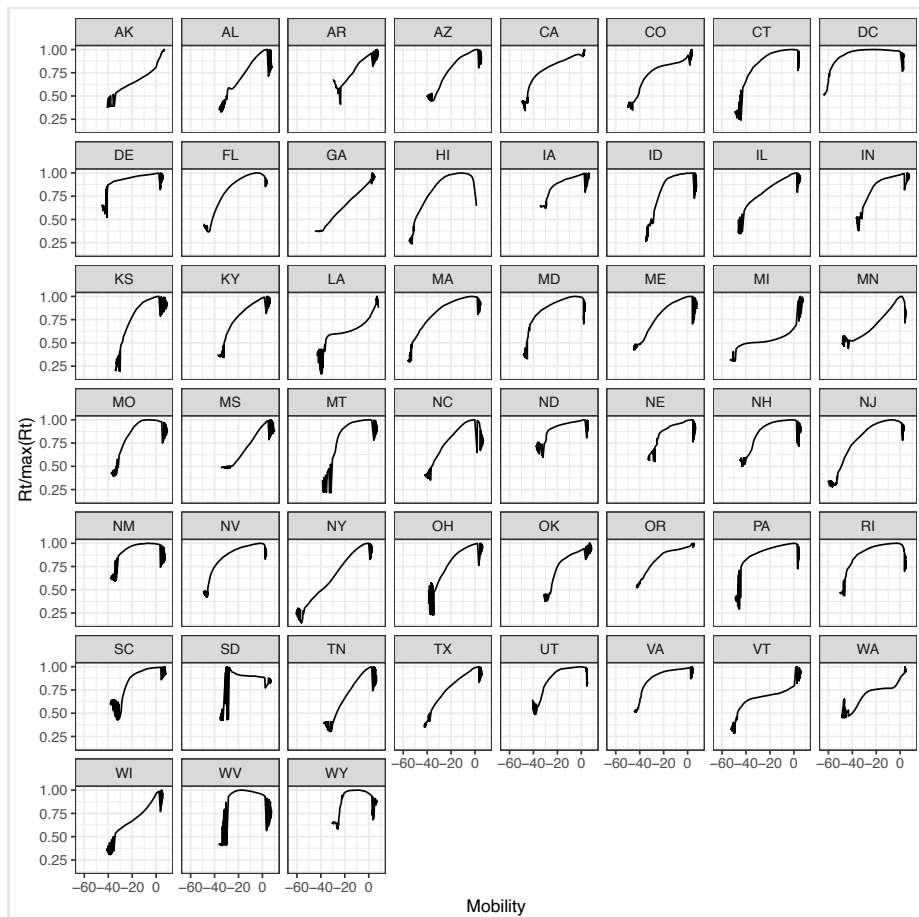

Figure 9: Relationship between mobility reductions (7 day moving average) and reductions in Rt by state.

**Scaling of testing across states**

The Bayesian model's upper bounding requires that access to testing be non-decreasing over time. While an estimate of the precise probability that an individual infected at a particular time will be diagnosed is difficult to measure, we can directly observe the massive increase in tests being done during the interval in which incidence increased dramatically. Figure 10 shows the total number of tests done per week in states with. For each state we see an approximatly monotonic increase in March, with increases continuing through April at a lower pace.

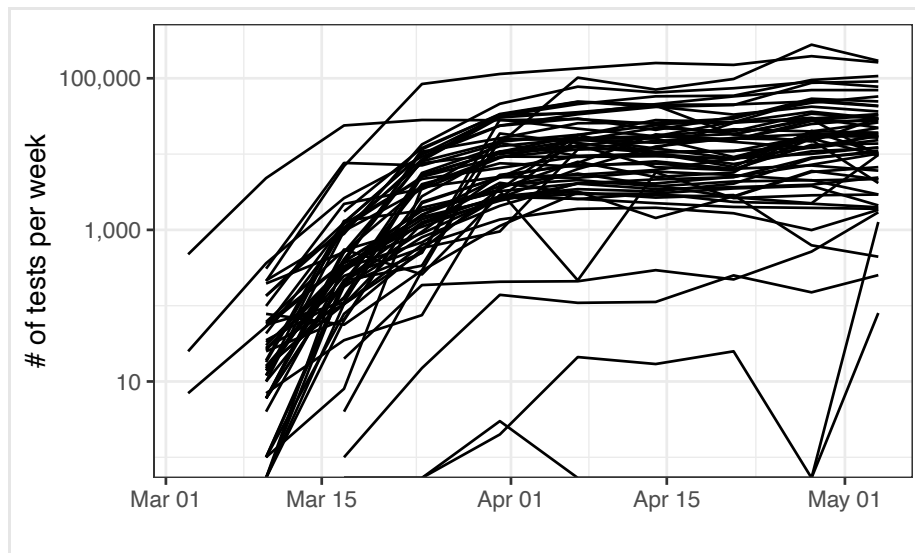

Figure 10: Weekly COVID-19 testing by state.

**Growth of uncertainty due to censoring**

When making public health decisions it is critical to utilize not just the point estimate of Rt or incidence, but also to take into account uncertainty around the estimate. Uncertainty in our estimates comes from two sources. The first is the stocasticity inherent in count data. The second and far more important contributor, is the uncertainty stemming from right censoring. Individuals infected recently are not observed in either the case or death counts and are essentaily censored from out view. This causes us to have limited information on recent time trends. Figure 11 shows how the standard deviation of the Rt estimate nearly quadruples on average from its low in early April and implies that case should be taken when interpreting recent parameter value estimates.

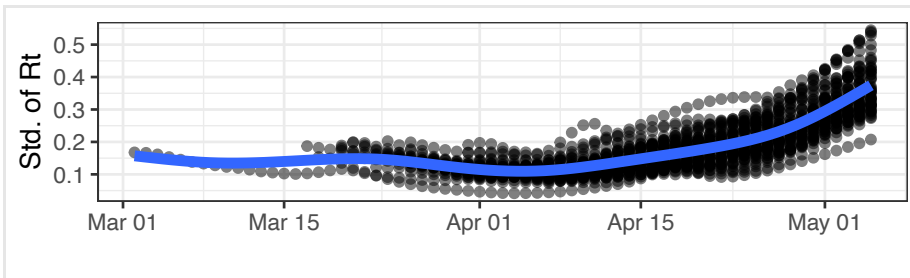

Figure 11: Standard deviation of the Rt estimate by date.

MAIN MANUSCRIPT FIGURES

Figure 1: US National estimated time varying reproductive rate (top). Incidence of fatal infections (bottom). Probability interval bands are shaded at the 50%, 90% and 95% levels.

Figure 2: Estimated effects of non-pharmaecutical interventions and national time trends on the time varying reproductive number (Rt) using a generalized additive model along with 95% bootstrap confidence intervals.

Figure 3: The relationship of time varying reproductive number (Rt), mobility (7 day moving average) derived from Google Community Mobility data, and time (7). States that never implemented a stay-at-home order are marked in green.

Figure 4: Trace plots for mobility (7-day moving average) and Rt by state. The thicker lines represent the mean values among states that have or have not yet enacted the NPI. The rug band at the bottom shows when the NPI were enacted.

Figure 1

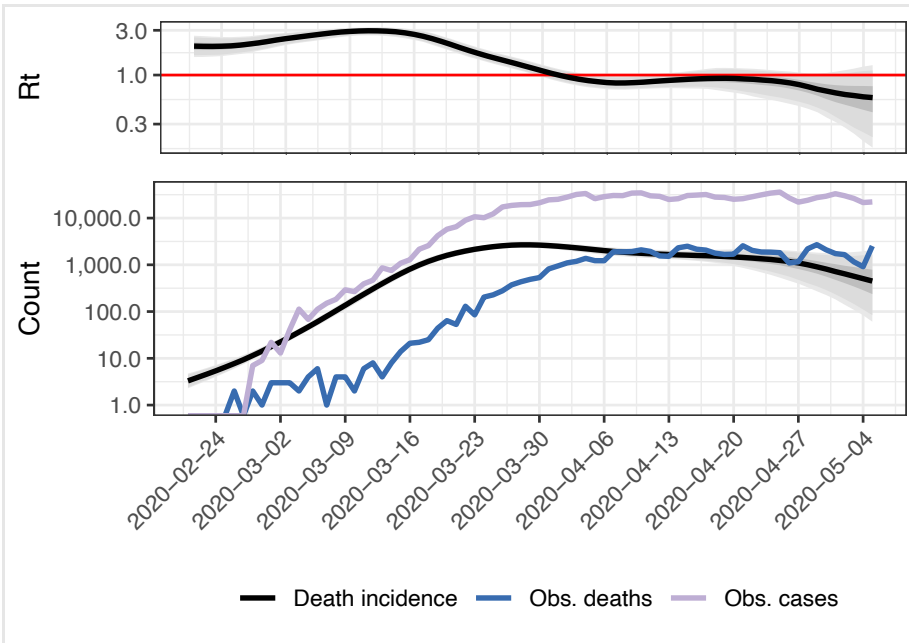

Figure 2:

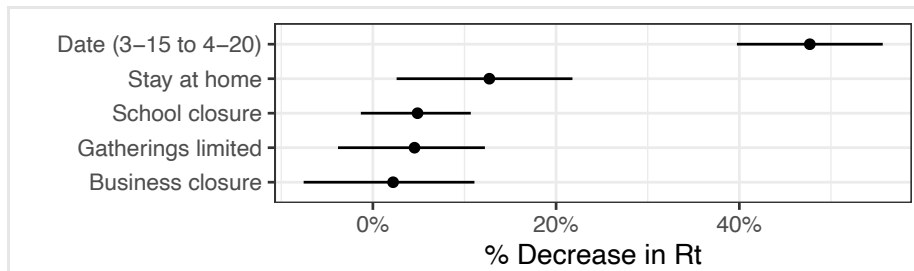

Figure 3:

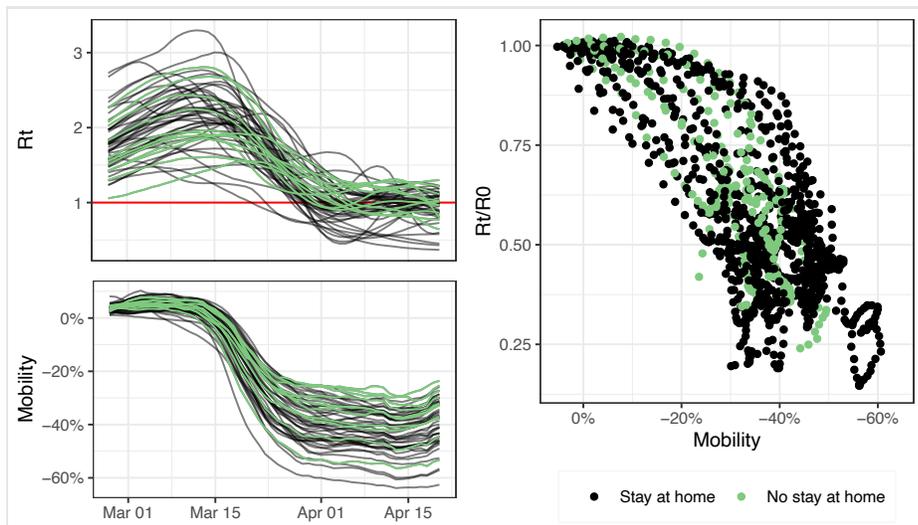

Figure 4:

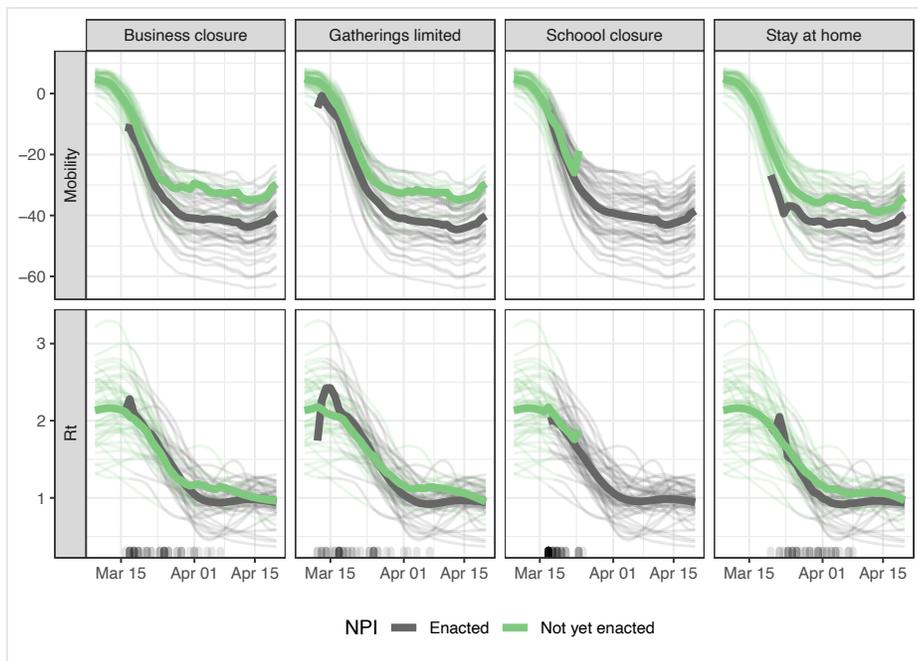